\documentclass[aps, longbibliography, amsmath,amssymb, superscriptaddress, 12pt]{revtex4-2}
 
\usepackage{amsmath}
\usepackage{amsfonts}
\usepackage{xcolor}
\usepackage{graphicx}
\usepackage{dcolumn}
\usepackage{bm}
\usepackage{hyperref}

\usepackage{nicematrix}
\usepackage{amsfonts}
\usepackage{amsthm}
\usepackage{amssymb}
\usepackage{float}
\usepackage{nicematrix}
\usepackage{tabularx}
\usepackage{physics}
\usepackage{siunitx}
\usepackage{bbold}
\usepackage[svgnames]{xcolor} 
\usepackage{xr}
\usepackage{soul}

\definecolor{eh}{rgb}{0.9, 0, 0}

\newcommand{\CVS}{$\mathrm{Cs}\mathrm{V}_3\mathrm{Sb}_5$}
\newcommand{\AVS}{$\mathrm{A}\mathrm{V}_3\mathrm{Sb}_5$}

\newcommand{\bk}{\mathbf{k}}
\newcommand{\bq}{\mathbf{q}}
\newcommand{\bQ}{\mathbf{Q}}

\newcommand{\qpiarc}[1]{\alpha_{#1}}
\newcommand{\qpiline}[1]{\lambda_{#1}}

\begin{document}


\title{Momentum Structure of Superconductivity and Sublattice Effects from Quasiparticle Interference in CsV$_3$Sb$_5$}

\author{\small Aaron~G.\ Greenberg}
\email[These authors contributed equally to this work.]{}
\affiliation{\footnotesize \mbox{Department of Physics, Yale University, New Haven, Connecticut 06520, USA}}
\affiliation{\footnotesize \mbox{Energy Sciences Institute, Yale University, West Haven, Connecticut 06516, USA}}

\author{\small Xinze Yang}
\email[These authors contributed equally to this work.]{}
\affiliation{\footnotesize \mbox{Department of Physics, Yale University, New Haven, Connecticut 06520, USA}}
\affiliation{\footnotesize \mbox{Energy Sciences Institute, Yale University, West Haven, Connecticut 06516, USA}}

\author{\small Junze Deng}
\email[These authors contributed equally to this work.]{}
\affiliation{\footnotesize \mbox{Department of Applied Physics, Aalto University School of Science, FI-00076 Aalto, Finland}}

\author{\small Pranab Kumar Nag}
\email[These authors contributed equally to this work.]{}
\affiliation{\footnotesize \mbox{Department of Physics, Yale University, New Haven, Connecticut 06520, USA}}
\affiliation{\footnotesize \mbox{Energy Sciences Institute, Yale University, West Haven, Connecticut 06516, USA}}

\author{\small Kirsty Scott}
\affiliation{\footnotesize \mbox{Department of Physics, Yale University, New Haven, Connecticut 06520, USA}}
\affiliation{\footnotesize \mbox{Energy Sciences Institute, Yale University, West Haven, Connecticut 06516, USA}}

\author{\small Yi Jiang}
\affiliation{\footnotesize \mbox{Donostia International Physics Center, P. Manuel de Lardizabal 4, 20018 Donostia-San Sebastian, Spain}}

\author{\small Haoyu Hu}
\affiliation{\footnotesize \mbox{Department of Physics, Princeton University, Princeton, New Jersey 08544, USA}}

\author{\small Chandra Shekhar}
\affiliation{\footnotesize \mbox{Max Planck Institute for Chemical Physics of Solids, 01187 Dresden, Germany}}

\author{\small Dong Chen}
\affiliation{\footnotesize \mbox{Max Planck Institute for Chemical Physics of Solids, 01187 Dresden, Germany}}
\affiliation{\footnotesize \mbox{College of Physics, Qingdao University, 266071 Qingdao, China}}

\author{\small Claudia Felser}
\affiliation{\footnotesize \mbox{Max Planck Institute for Chemical Physics of Solids, 01187 Dresden, Germany}}

\author{\small Santiago Blanco-Canosa}
\affiliation{\footnotesize \mbox{Donostia International Physics Center, P. Manuel de Lardizabal 4, 20018 Donostia-San Sebastian, Spain}}
\affiliation{\footnotesize \mbox{IKERBASQUE, Basque Foundation for Science, Bilbao, Spain}}

\author{\small P\"{a}ivi T\"{o}rm\"{a}}
\affiliation{\footnotesize \mbox{Department of Applied Physics, Aalto University School of Science, FI-00076 Aalto, Finland}}

\author{\small B. Andrei Bernevig}
\affiliation{\footnotesize \mbox{Department of Physics, Princeton University, Princeton, New Jersey 08544, USA}}
\affiliation{\footnotesize \mbox{Donostia International Physics Center, P. Manuel de Lardizabal 4, 20018 Donostia-San Sebastian, Spain}}
\affiliation{\footnotesize \mbox{IKERBASQUE, Basque Foundation for Science, Bilbao, Spain}}

\author{\small Eduardo H.\ da Silva Neto}
\email[Corresponding Author: ]{eduardo.dasilvaneto@yale.edu}
\affiliation{\footnotesize \mbox{Department of Physics, Yale University, New Haven, Connecticut 06520, USA}}
\affiliation{\footnotesize \mbox{Energy Sciences Institute, Yale University, West Haven, Connecticut 06516, USA}}
\affiliation{\footnotesize \mbox{Department of Applied Physics, Yale University, New Haven, Connecticut 06520, USA}}

\maketitle

\vspace{-3mm}

{\small Quantum interference encoded in the sublattice texture of kagome Bloch wavefunctions has been widely invoked as a route to correlated states, including chiral charge order, unconventional superconductivity, and their possible intertwining in pair-density-wave (PDW) states. Using sub-Kelvin scanning tunneling microscopy, we conducted spectroscopic mapping of the kagome material \CVS{} with high energy resolution and dense energy sampling through the superconducting gap. Quasiparticle interference (QPI) analysis, aided by \textit{ab initio} and symmetry calculations, reveals an isotropic superconducting gap on the Fermi surfaces derived from V $M_z$-even ($M_z^+$) $d$ orbitals, thereby constraining possible gap symmetries and limiting any gap anisotropy to the remaining V $M_z$-odd ($M_z^-$) and Sb $p_z$ bands. Meanwhile, the CDW-peak-selected $\mathrm{d}I/\mathrm{d}V$ spectra closely track the spatially averaged density of states and show no distinct enhancement restricted to subgap energies, which do not support an additional PDW modulation within our sensitivity.
Finally, the selective absence of specific QPI scattering vectors points to a spectroscopic sensitivity to sublattice character on the Fermi surface. Together, these results provide a clearer experimental picture of the low‑energy electronic structure relevant to kagome superconductivity in \CVS{}.

\clearpage

\vspace{3mm}
\section*{Introduction}

The kagome lattice, consisting of three hexagonal sublattices that form corner-sharing triangles, features an electronic structure with flat bands, van Hove singularities, and Dirac points \cite{Guo2009, Kiesel2012, Kiesel2013, Kang2022}, making it a fertile platform for exploring diverse electronic phenomena. 
The electron wavefunctions in momentum space are characterized not only by momentum and orbital characters, but also by their matrix element amplitudes and phases on different sublattices \cite{Hu2022RichNature, Brouet_CoSn_sublattice_interference}. Theoretically, this sublattice character plays a pivotal role in enabling quantum interference effects that produce flat bands and symmetry-enforced band crossings, enhancing susceptibility to correlation-driven phases such as magnetism, charge density waves (CDWs), and unconventional superconductivity \cite{Checkelsky2024}.

Despite the importance attributed to sublattice character in kagome materials, the sublattice composition has not been experimentally probed at the Fermi level ($E_F$). Prior studies have generally inferred sublattice character either because the band dispersions match angle resolved photoemission spectroscopy (ARPES), or has relied on orbital analyses at binding energies hundreds of meV below $E_F$ and then assigned sublattice character based on that orbital content \cite{Hu2022RichNature, Lan2026Sublattice}. This does not probe the actual sublattice composition at $E_F$, where kagome derived bands can mix orbitals strongly and where the phases of the matrix elements control the interference pattern. Because the relevant instabilities and ordered phases are governed by the structure of the Fermi surface, characterizing the sublattice contributions at $E_F$ can provide valuable context for assessing the role of kagome‑related interference effects in emergent phenomena.

The appearance of superconductivity in the \AVS{} class of kagome materials, with $T_c$ ranging from $0.9$ K to $2.5$ K, also opened a potential avenue for the investigation of kagome-related effects on superconductivity \cite{Ortiz2020, KVSTc, RVSTc, Roton, Threedim, AVSRev, SmallFermiPockets, Deng2024}.
In \CVS{}, the material with the highest $T_c$ among its stoichiometric counterparts \cite{Ortiz2020, AVSRev}, the existence of multiple CDW instabilities \cite{Cascade}, together with reports of electronic nematicity and time-reversal symmetry breaking, suggests a complex electronic structure characteristic of unconventional superconductivity \cite{ Kivelson_Intertwined_2015}. 
In this landscape of nearly degenerate broken-symmetry phases, it is natural to hypothesize that superconductivity is intertwined with other orders and their fluctuations, akin to many unconventional superconductors \cite{Keimer2015, fernandes2022iron, Nuckolls_Yazdani_2024}. Signatures of this intertwining may include a highly momentum-dependent superconducting gap or spatially modulated superconductivity with periodicity related to the CDW, often referred to as a pair-density wave (PDW) state.
Interestingly, reports using different experimental probes have indicated various scenarios for the superconducting gap structure, ranging from isotropic to nodal gaps, and even their coexistence \cite{Mu_2021_NMR, Xu2021, Roppongi2023, Hu2024_twogaps, Hossain2025}. However, many of these measurements do not directly measure the momentum structure of the gap. Meanwhile, while angle-resolved photoemission spectroscopy measurements (ARPES) can, in principle, be employed to resolve this issue, the small $T_c$ and size of the superconducting gap pose significant challenges \cite{mine2024directobservationanisotropiccooper}. 
Consequently, measurements capable of resolving the superconducting gap structure in momentum space are not only desirable, but crucial for understanding kagome superconductivity.

Scanning tunneling spectroscopy (STS) is an ideal method for investigating superconductivity in \CVS{}, as it can resolve the momentum structure of the superconducting gap through quasiparticle interference (QPI) signals with high energy resolution at sub-Kelvin temperatures \cite{mcelroy2003octet, Sprau_Science_orb_selective_cooper, Nag2024a}. It is also able to simultaneously probe both the superconducting and CDW states, thus testing their relationship.
Equally important in a multi‑sublattice system, QPI intensities carry the matrix‑element amplitudes and phases that are sensitive to sublattice weights, and the selective appearance or absence of specific scattering vectors encodes the underlying Fermi‑level sublattice character.
Previous STS measurements of \CVS{} showed U-shaped tunneling gaps, consistent with conventional \textit{s}-wave superconductivity, that sometimes feature a subtle two-gap structure in the tunneling density of states (DOS), suggesting distinct gap structures on different Fermi surfaces \cite{Xu2021, Hu2024_twogaps}. However, the detailed shape of the tunneling spectra and the presence of a clear two-gap structure varied between measurements, possibly due to local strain or tip-related effects on the tunneling matrix element \cite{Xu2021, Hu2024_twogaps}. Additionally, there have been reports of a spatially modulated superconducting gap, interpreted as an unconventional pair-density wave state \cite{Roton, Deng2024}.
Besides these efforts, \textit{spectroscopic‑imaging} experiments using QPI to resolve the momentum structure of the superconducting gap remain limited.  Reported experiments sampled only a few energies near the Fermi level, which is insufficient for determining the momentum structure of the superconducting gap \cite{Sun2024_low_sampling}. 
This task is further complicated by the presence of multiple bands arising from the V-based kagome lattice in \AVS{} \cite{Hu2022RichNature, Kang2022}, making it difficult to identify which parts of the Fermi surface contribute to the signals observed in QPI signals. A systematic study with high energy-resolution and dense sampling, together with a detailed mapping between band structure and QPI features, is therefore essential.

Here, we report sub-Kelvin STS measurements of \CVS{} with the requisite energy resolution and dense energy sampling, combined with \textit{ab initio} calculations of the electronic structure as observed through STS. 
The \textit{ab initio} calculations allow us to establish a Fermi‑surface‑to‑QPI mapping, which we refer to as the $\bk\to\bq$ mapping, showing that the QPI response is sensitive to the momentum dependence of the superconducting gap on the Fermi surface centered at the \textit{K} point, derived from V \textit{d}‑orbitals that are even under mirror reflection $M_z$ ($M_z^+$ band).
Analysis of the experimental QPI also reveals the selective absence of certain features at $\mathbf{q}$ locations where the $\mathbf{k}$ states connected by $\mathbf{q}$ exhibit reduced overlap in sublattice weight.
With the $\mathbf{k}\rightarrow\mathbf{q}$ mapping established, we determine that the superconducting gap structure in the $M_z^+$ kagome-derived Fermi pockets is isotropic. In addition, the CDW‑peak‑selected spectra closely track the spatially averaged density of states and show no distinct enhancement at subgap energies, and therefore do not offer particularly strong support for a PDW state.

\vspace{3mm}
\section*{Results}
\noindent \textbf{{Surface termination preparation}}

Figure \ref{Fig1}\textbf{a} shows the crystal structure of \CVS. A natural cleaving plane exists between the weakly bonded Cs and Sb layers, resulting in surface terminations that range from Cs-rich surfaces with scattered vacancies to Sb-terminated surfaces with a few residual Cs atoms.
Direct tunneling access to the kagome V termination may be desirable, but the bond between Sb and V is not broken during the cleaving process, meaning that the Sb layer remains the surface closest to the kagome structure. 
Although the Sb termination is relatively large, it is always accompanied by scattered residual Cs atoms, and the tunneling junction directly on these sporadic Cs sites is highly unstable compared to tunneling on the fully formed Sb layer.

Following previous works, we use the STM tip to \textit{sweep} away residual Cs atoms, preparing a Cs-free Sb surface for experiments \cite{Roton, Xing2024}. This process is illustrated in Figs.\,\ref{Fig1}\textbf{b}–\textbf{e}, where the motion of Cs atoms appears as zigzag patterns in the STM images due to their continuous displacement during topographic scans. The resulting Cs-free regions reveal the underlying Sb atoms and the CDW features, Figs.\,\ref{Fig1}\textbf{f},\textbf{g}. 
This results in a \CVS-specific issue that has received limited attention: repeated Cs–tip interactions during the \textit{sweeping} procedure can modify the tip apex and introduce artifacts in the tunneling conductance. 
Our measurements were verified using a different tip–sample junction and a separate nanoscopic area on the sample, and the salient features were reproduced in both cases (Supplementary Materials Section \ref{sup-SM:Locations}).

\vspace{3mm}
\noindent \textbf{{STS measurements near the Fermi level}}

To investigate superconductivity, we performed STS imaging measurements focusing on the energy range of the superconducting gap ($\pm 0.5$\,meV).
Performing measurements at a base temperature of $0.5$\,K both in the superconducting state ($B=0$\ T) and with superconductivity suppressed by a magnetic field ($B=3$\ T), we are able to isolate the effects of superconductivity on STS, while preserving maximum energy resolution with low temperature.
Figures \ref{Fig2}\textbf{a}–\textbf{e} and \ref{Fig2}\textbf{f}–\textbf{j} show the Fourier transforms of STS images, $g(\bq,E)$, at selected energies from datasets measured between $-1.5$ and $1.5$\,meV in $50$\,$\mu$eV steps, at $B=0$\,T and $B=3$\,T respectively. To the best of our knowledge, these represent the most finely energy-sampled spectroscopic mapping measurements of the superconducting gap in \CVS{} reported to date.

The $g(\bq,E)$ reveals two distinct types of features: sharp peaks mostly corresponding to various CDW modulations, and extended arcs and lines associated with QPI, Figs.\,\ref{Fig2}\textbf{a}–\textbf{j}. These features can be further highlighted by integrating $g(\bq,E)$ over $E$ in the $-1.5$ to $1.5$\ meV range, Fig.\ \ref{Fig2}\textbf{l}. The dashed hexagon outlines the scattering Brillouin zone, where six Bragg peaks, $\bQ_{\textrm{Bragg}}$, associated with the Sb lattice are visible as peaks at the corners of the hexagon. The CDW peaks ($\bQ_{1/2}$) marked by purple circles correspond to the 2 $\times$ 2 CDW. In addition, a peak corresponding to the 1 $\times$ 4 CDW ($\bQ_{1/4}$) is present along the $q_x$ axis (orange circles).
A related peak also appears along the same direction at $|\bq| = |\bQ_{\textrm{Bragg}}| - |\bQ_{1/4}|$ (brown circles).
Peaks with similar $|\bq|$ along the other two lattice directions are also observed, denoted $\bQ^*_{3/4}$ in Fig.\ \ref{Fig2}\textbf{m}.
Previous reports note that while $\bQ_{3/4}$ remains essentially nondispersive, the $\bQ_{3/4}^*$ may show a weakly energy-dependent $|\bQ^*_{3/4}|$ \cite{SmallFermiPockets}, suggesting a different origin for the two features. In our data, we cannot exclude the possibility that the two features differ, and thus we analyze $\bQ_{3/4}$ and $\bQ_{3/4}^*$ separately.
Figure~\ref{Fig2}\textbf{m} schematically depicts the locations of these various peaks. Additionally, extended QPI features are observed, with arc-like structures labeled $\qpiarc{}$ and line-like structures labeled $\qpiline{}$ in Fig.\ \ref{Fig2}\textbf{n}. 
{Consistent with previous reports \cite{Cascade, Roton}, the circular QPI feature associated with the Sb $p_z$ band is weak near the Fermi level, but becomes more apparent in larger-bias topographs, which integrate the electronic structure over a broader energy window, as seen in Fig.\,\ref{Fig1}\textbf{g}.} 

An anisotropic superconducting gap manifests itself in QPI measurements in two primary ways. First, it modifies the contours of constant energy (CCEs) within the superconducting state, producing smaller, geometrically distinct Bogoliubov pockets that generate additional QPI wave vectors at $\bQ$ locations where no QPI intensity appears in the normal (\textit{i.e.}, non-superconducting) state \cite{mcelroy2003octet, Sprau_Science_orb_selective_cooper, Nag2024a}. The interference patterns associated with these quasiparticles are commonly referred to as Bogoliubov QPI (BQPI).
Second, the intensity of normal‑state QPI features, which map to specific regions of the Fermi surface, should show an energy‑dependent suppression reflecting the superconducting gap at the associated $\bk$ points. In the presence of gap anisotropy, different QPI features would be expected to lose intensity at different energies \cite{Nag2024a}, effectively revealing distinct gap energy scales for different $\bq$ vectors.
Close inspection of $g(\bq,E)$ (Fig.\,\ref{Fig2} and Supplementary Fig.\,\ref{sup-fig:All-Energies-0T}) shows no clear evidence of BQPI, which would have represented the first type of indicator of gap anisotropy. Our analysis consequently focuses on the energy‑dependent intensities at those $\bq$ locations where QPI intensity is already present in the normal state.

\vspace{3mm}
\noindent \textbf{{$\mathbf{k} \rightarrow \mathbf{q}$ mapping and sublattice effects}}

A critical step in investigating the superconducting gap structure through QPI measurements is obtaining a precise $\mathbf{k} \rightarrow \mathbf{q}$ mapping. To this end, we first compute the surface-projected Fermi surface on the STM-measured (Sb-terminated) surface using \textit{ab initio} methods, yielding the Fermi surface in Fig.~\ref{Fig3}\textbf{a}. To account for the twofold-symmetric QPI pattern induced by the $\bQ_{1/4}$ stripe order (\textit{e.g.}, Fig.~\ref{Fig2}\textbf{l}), we include $\bQ_{1/4}$ folding by adding a $4\times1$ periodic on-site potential on the V kagome lattice [for details see \textit{CDW and unfolding} in Methods], which yields the calculated QPI and joint density of states (JDOS) patterns in Figs.~\ref{Fig3}\textbf{b} and \textbf{c}. 
The resulting Fermi surface, consistent with ARPES measurements \cite{Kang2022}, can be broadly categorized into three components. (\textit{i}) Small triangular pockets centered at the $\bar{K}$ point, derived from V \textit{d}-orbitals that are even under mirror reflection $M_z$ ($d_{x^2 - y^2}$, $d_{xy}$, and $d_{z^2}$; denoted as $M_{z}^+$ orbitals). (\textit{ii}) Larger triangular contours, also centered at $\bar{K}$, derived from $M_z$-odd orbitals ($d_{xz}$ and $d_{yz}$; denoted as $M_z^{-}$ orbitals). (\textit{iii}) A quasi-circular contour centered at $\bar{\Gamma}$, derived from Sb $p_z$ orbitals, which does not contribute to the low-energy QPI features in our measurement (for details see Supplementary Section \ref{sup-app:qpi-sb-pz}).

The question then arises: how does one uniquely identify each QPI pattern, $\alpha$ and $\lambda$, with specific $\mathbf{k}$ points in the band structure? To address this, we begin by noting that a Green's function calculation restricted to the V $M_z^+$ bands reproduces the experimentally observed $\alpha$ and $\lambda$ features (Fig.\,\ref{Fig3}\textbf{b}). A closer examination of the \textit{ab initio} results and their comparison to the experimental data clarifies why the $M_z^+$ sector is the origin of the $\alpha$ and $\lambda$ features. QPI from the Sb $p_z$-derived Fermi surface is excluded as the circular QPI feature characteristic of this band is absent at the low energies probed in our experiments (Supplementary Section \ref{sup-app:qpi-sb-pz}). The $M_z^-$ sector is likewise inconsistent with both the geometry and dispersion of the $\alpha$ and $\lambda$ features, and the $\mathbf{Q}_{1/4}$ CDW folding further suppresses the $M_z^-$ Fermi surface (for details see Supplementary Section \ref{sup-app:qpi-mz-odd-gap}). 
Thus, the experimentally observed QPI features arise entirely from the $M_z^+$ bands.

The \textit{ab initio} calculations also provide the full sublattice‑resolved wavefunctions on the $M_z^+$ Fermi surface, enabling us to further examine how sublattice character influences the QPI response. To isolate this effect, we compare the QPI results with a JDOS calculation based solely on the $M_z^+$ Fermi‑surface DOS (Fig.\,\ref{Fig3}\textbf{c}), which does not incorporate sublattice information. The JDOS predicts additional features, labeled $\delta_{1\text{--}3}$, that do not appear in the experimental data (purple dashed lines in Fig.\,\ref{Fig3}\textbf{c}). To understand this discrepancy, we analyze the relevant scattering vectors directly on the \textit{ab initio} Fermi surface. Figures \ref{Fig3}\textbf{d--f} show the $M_z^{+}$ contours, forming broken, rounded triangles composed of arcs and nearly horizontal segments, and color-coded by sublattice weight. Scattering processes connecting states with nearly parallel group velocities are strongly suppressed, an effect naturally captured in Green's-function calculations of QPI~\cite{Yang2025}, and this removes some of the JDOS-predicted features (dashed vectors in Fig.\,\ref{Fig3}\textbf{d--f}). However, even after imposing this constraint, the $\delta_{1\text{--}3}$ processes would still be allowed. Figure \ref{Fig3}\textbf{f} further illustrates that the $\bk$ states involved in $\delta_{1\text{--}3}$ have weaker sublattice-weight overlap than those producing the $\alpha$ and $\lambda$ features.
This reduced overlap should strongly diminish the intensity of the $\delta$ features, offering a natural explanation for their absence in the experimental data.

The full situation is, of course, more intricate. The sublattice weights vary smoothly and overlap across $\mathbf{k}$‑space, and the complete wavefunctions include not only sublattice but also orbital components. These ingredients are incorporated in the QPI simulations based directly on the \textit{ab initio} wavefunctions (Fig.\,\ref{Fig3}\textbf{b}), which naturally reproduce, for example, the suppression of the $\delta_2$ structure relative to $\alpha$ and $\lambda$ features, as well as the discontinuity between the $\alpha_1$ and $\lambda_1$ features, both of which are consistent with the experimental data. 
{A fully quantitative comparison with experiment would require detailed knowledge of the scattering potential, which determines the relative weights of QPI features across $\mathbf{q}$-space and lies beyond the scope of our modeling. Nevertheless, the overall agreement suggests that sublattice character is a major factor in determining which QPI features appear in the data.}
These considerations also highlight why simplified tight‑binding models may overlook essential aspects of the real electronic states in kagome materials.

\vspace{3mm}

\noindent \textbf{{Momentum Resolved Gap Structure}}

The $\bk \rightarrow \bq$ mapping establishes that the $\qpiarc{}$ features correspond to the rounded corners of the quasi-triangular $M_z^+$ pockets, while the $\qpiline{}$ features trace their flat sides. To quantify their energy dependence, we use our finely energy-sampled spectroscopic maps $g(\bq,E)$, applying at each energy masks (Fig.\,\ref{Fig4}\textbf{a}) that isolate the regions of $\bq$ space associated with $\qpiarc{1-3}$ and $\qpiline{1-3}$. For each mask and each energy, we average $g(\bq,E)$ over the masked region, yielding energy spectra for the $\alpha$ and $\lambda$ features. Figures \ref{Fig4}\textbf{b} and \textbf{c} show these spectra, each exhibiting a superconducting gap. Since the individual arcs or lines within their respective groups are symmetry-equivalent, Fig.\ \ref{Fig4}\textbf{d} presents the spectra obtained by averaging the three arc spectra in Fig.\,\ref{Fig4}\textbf{b} (the $\alpha$ curve) and the three line spectra in Fig.\,\ref{Fig4}\textbf{c} (the $\lambda$ curve), along with the real-space-averaged spectrum (obtained by averaging over the field of view). Remarkably, all $\alpha$ and $\lambda$ spectra exhibit superconducting gaps identical to that of the real-space-averaged spectrum, demonstrating that the $M_z^+$ pockets host an isotropic superconducting gap.
{Based on our experimental parameters, including temperature and lock-in modulation, together with the analysis of different segments of the arc and line features, we estimate that any residual anisotropy of the superconducting gap on the $M_z^+$ pockets is smaller than $\pm 6\%$ (Supplementary Section \ref{sup-SM:QPI_Anisotropy_Sensitivity}).}

\vspace{3mm}
\noindent \textbf{{Relation between CDW and superconductivity}} 

A PDW state produces a spatially modulated superconducting gap, that should appear in STS as periodic changes in the gap amplitude. 
When the modulation period of the superconducting gap coincides with that of a preexisting CDW, {or with another periodic structure such as the atomic lattice}, caution is required, as modulations of the underlying electronic DOS can lead to apparent, but trivial, variations of the gap. 
{Such effects can appear in one-dimensional line cuts through a limited set of atoms, where the extracted gap may show small variations, on the order of a few percent, with the same periodicity as the lattice or CDW. Our data also show this effect (Supplementary Section \ref{sup-SM:STS_Real_Space_Sensitivity}), demonstrating the sensitivity of our measurements to small spectral variations. While such variations could in principle arise from a PDW, a more robust PDW signature would be either a modulation at a wave vector $\bQ$ absent in the normal state \cite{Zhao2023_PDW}, or a modulation at an existing normal-state wave vector whose energy dependence differs significantly from the spatially averaged superconducting spectrum. We therefore examine whether any CDW-associated two-dimensional Fourier peaks, which probe many more atoms than a one-dimensional line cut, exhibit gap-specific deviations from the spatially averaged superconducting spectrum.}

In prior work on \CVS{}, the $\bQ^*_{3/4}$ and ${\bQ}_{3/4}$ modulations were grouped together and interpreted as signatures of a single PDW state \cite{Roton}. It was further suggested that these peaks become more prominent within the superconducting gap, although a continuous, densely sampled energy‑dependent measurement of these features has not been reported. A PDW has also been associated with the $\bQ_{1/2}$ modulation in subsequent studies \cite{Deng2024}.
In contrast, our data show that the spectra associated with these modulations largely track the spatially averaged DOS: they exhibit substantial amplitude across the full $\pm 10$ meV energy range and show no salient deviations within the superconducting gap (Figs.\,\ref{Fig4}\textbf{e--f}). Likewise, the high‑energy‑resolution spectra in the $-1$ to $1$ meV window, for both the superconducting ($B=0$\,T) and normal ($B=3$\,T) states, also follow the spatially averaged DOS (Figs.\,\ref{Fig4}\textbf{g}).

\section*{Discussion}

We used sub‑Kelvin scanning tunneling spectroscopy (STS) with high energy resolution and dense sampling to search for signatures of unconventional superconductivity in the kagome superconductor \CVS{}. In doing so, we used realistic \textit{ab initio} calculations to establish a robust $\mathbf{k}\rightarrow\mathbf{q}$ mapping and to examine how QPI encodes sublattice interference effects on the Fermi surface. Such sensitivity, particularly when probing the Fermi surface directly, may be relevant for understanding interaction effects and potential unconventional superconductivity in kagome metals \cite{Kiesel2012, Kiesel2013, Wu2021}. The $\mathbf{k}\rightarrow\mathbf{q}$ mapping further enables a direct determination of the momentum dependence of the superconducting gap from the QPI signal. Regarding a possible PDW state, we do not find clear evidence of superconducting gap modulations beyond those expected when superconductivity develops on top of preexisting DOS modulations. 
{It remains possible that the superconducting order consists of a mixture of uniform and PDW components, with the former dominating and any PDW component remaining below our experimental sensitivity.}
In this scenario, our measurements suggest that if a robust PDW signature exists in STM data, future investigations should focus on the $\bQ_{3/4}$ modulation, which exhibits a small deviation from the average DOS, reflected as a kink within the superconducting gap (Fig.\,\ref{Fig4}\textbf{e} and Supplementary Fig.\,\ref{sup-fig:Location-Comparison}). Interestingly, a similar feature is not observed in the related $\bQ_{1/4}$ spectrum, although this discrepancy could arise from distinct coupling strengths between individual CDW components and superconductivity. Nevertheless, we remain cautious in interpreting this anomaly as evidence of a PDW state, since, consistent with previous STS studies that showed two-gap structures in the $dI/dV$ spectra \cite{Hu2024_twogaps}, the relative strength of this spectral feature in the $\bQ_{3/4}$ Fourier peak varies with the measurement location on the sample and with changes in the tip–sample tunneling junction (Supplementary Fig.\,\ref{sup-fig:Location-Comparison}).
Relatedly, recent theory shows that an enhanced PDW pairing susceptibility in kagome models does not always imply the stabilization of a superconducting PDW phase~\cite{Lamponen2025SuperconductivityA}.

The QPI signal, with the proper $\mathbf{k}\rightarrow\mathbf{q}$ mapping, reveals an isotropic superconducting gap on the V $M_z^+$ Fermi surfaces. 
{Based on general symmetry considerations (\textit{e.g.}, see Ref.\,\cite{Holbaek2023}), the observed isotropy of the V $M_z^+$ gap rules out several symmetry-allowed gap structures that would produce different gap magnitudes along the measured arcs and line segments of the $M_z^+$ Fermi surface, thereby narrowing the set of possible superconducting gap symmetries.}
{Furthermore, given our measurement of the superconducting gap on the V $M_z^+$ band, any reported gap anisotropy in \CVS{} \cite{Hossain2025, Roppongi2023} would have to arise from other bands, such as the $M_z^-$ or Sb $p_z$ bands.}
In connection with specific theoretical models, a recent proposal suggests that an $s_{\pm}$ gap structure, with nearly isotropic gaps of opposite signs on different Fermi surfaces, can emerge from fluctuating chiral CDW  order \cite{schultz2025superconductivitykagomemetalssoft}. However, in that framework, if such CDW fluctuations exist, they would couple primarily to the $M_z^-$ or Sb-derived Fermi surfaces, and the gap on the V $M_z^+$ surface was not explicitly computed. 
First‑principles studies have also examined CDW physics and superconductivity in \CVS{}, identifying phonons as pairing mediators \cite{Gutierrez-Amigo2024PhononA,Alkorta2025Symmetry-brokenA}.
The results reported here provide cornerstones for establishing a comprehensive theory, grounded in realistic electronic structure, that captures both the CDW and superconducting orders within a unified and physically transparent framework.

\section*{Data availability}
The data that support the findings of this study are available from the corresponding authors
upon request.

\bibliography{CVS}

\section*{Methods}
\noindent \textbf{Sample Growth and STM Experiment}\\
The growth of high-quality single crystals of \CVS was achieved by employing the self-flux method, as previously outlined in Ref. \cite{Ortiz2020}. Initial elements of high purity were loaded into an alumina crucible, which was then sealed in a tantalum tube and subsequently in a quartz tube. The loaded quartz tube was heated to 1000 $^\circ$C and maintained for 20 hours in a box furnace to ensure the homogenization of the entire liquid. The furnace was then cooled to 400 $^\circ$C at a rate of 3 $^\circ$C/h, after which it was switched off. The removal of the flux was achieved through its dissolution in water. The crystals are hexagonal thin plates and are air stable. The X-ray diffraction pattern of a plate-like crystal manifests as a series of peaks at the (00$\ell$) reflection, which indicates the orientation of as-grown crystals. Temperature dependent resistivity and magnetization measurements manifest a charge density wave transition at $\approx$ 94 K.
Samples are mounted on copper shims using two-part conducting silver epoxy (Epotek H20E) and a cleaving post is glued on top with conducting silver epoxy. 
Measurements are performed in a customized Unisoku USM-1300 STM. The samples were cleaved \textit{in situ} in an ultra-high vacuum environment with pressure below $5 \times 10^{-10}$ Torr at a temperature of $\approx80$\,K before being immediately inserted into the STM head at $4.2$\,K. For measurements performed at the lowest temperature of the STM instrument, the temperature sensor located at the STM head read $298$\,mK. At that thermometer temperature, the electronic temperature was estimated to be $520$\,mK based on Dynes fits to measurements of the superconducting gap on Pb. Refer to \cite{Nag2024a} for further details.\\

\noindent \textbf{STS Parameters} \\
Differential conductance measurements were acquired with a set point bias and current of -10 mV and 700 pA respectively over a 256 \(\times\) 256 grid on a 33 \(\times\) 33 \(\mathrm{nm^2}\) area. For all spectroscopic maps {with energy range between -1.5 and 1.5 meV}, the lock-in bias modulation was 50 \(\mathrm{\mu V}\) at 953 Hz. {For the spectroscopic map with energy range -9.6 to 9.6 meV, the lock-in bias modulation was 100 \(\mathrm{\mu V}\) at 953 Hz.} Details of the data analysis are described in the Supplementary Sec.\,\ref{sup-SM:Data_analysis}. \\

\noindent \textbf{QPI Calculation through First-principles Method}
\paragraph{First-principles calculation.} 
First-principles calculations were carried out within density functional theory (DFT) using the projector augmented-wave (PAW) method~\cite{Blochl1994ProjectorA,Kresse1999FromA}, as implemented in the Vienna \textit{ab initio} simulation package (VASP)~\cite{Kresse1996EfficiencyA,Kresse1996EfficientA}. The exchange-correlation potential was treated within the generalized gradient approximation (GGA) using the Perdew–Burke–Ernzerhof (PBE) functional~\cite{Perdew1996GeneralizedA}. A plane-wave energy cutoff of 500~eV was used, and the Brillouin zone (BZ) was sampled with a $\Gamma$-centered Monkhorst–Pack $\bk$-point grid of $12 \times 12 \times 9$~\cite{Monkhorst1976SpecialA}. Irreducible representations of the electronic states were obtained using \texttt{IRVSP}~\cite{Gao2021IrvspA}. Maximally localized Wannier functions (MLWFs) were constructed for Cs-$s$, V-$d$, and Sb-$p$ orbitals~\cite{Mostofi2008Wannier90A,Mostofi2014An-updatedA,Pizzi2020Wannier90A}. Based on these MLWFs, the surface spectral function on the Sb-terminated surface can be obtained by the iterative Green’s function method \cite{Sancho1984QuickA, Sancho1985HighlyA} as implemented in WannierTools~\cite{Wu2018WannierToolsA}.

\paragraph{QPI simulation}
The QPI in momentum space was evaluated from the retarded Green's function,
\begin{equation}
    G(\vb{q}, \omega) = \sum_{\bk} \bqty{ G_0(\bk, \omega) + G_0(\bk, \omega)\, T(\omega)\, G_0(\bk+\vb{q}, \omega) },
    \label{eqn:methods-qpi}
\end{equation}
where the $T$ matrix is given by
\begin{equation}
    T(\omega) = V_{\bk} \bqty{\mathbb{1} - V_{\bk} \sum_{\bk} G_0(\bk, \omega)/N }^{-1},
\end{equation}
with $N$ the number of sampled $\bk$ points. Impurities are assumed to be nonmagnetic and randomly distributed, so the scattering potential is approximated as $V_{\bk} = V_0 \mathbb{1}$, with $\mathbb{1}$ the identity matrix and $V_0=50$~meV, a constant on-site potential at the defect site. The QPI spectra were then computed using the convolution technique in Refs.\;\cite{Capriotti2003Wave-vectorA,Wang2003QuasiparticleA,Fang2013TheoryA,Marques2021TomographicA} on a $400\times 400$ $\vb{q}$-mesh covering the entire surface Brillouin zone.

\paragraph{CDW and unfolding.}
The $4\times 1$ stripe order observed on the Sb-terminated surface~\cite{Cascade,Ortiz2021FermiA,Hu2022RobustnessA} was modeled by adding a periodic on-site potential to the V kagome layer
$V_{\beta} = V_0 \cos(2 \pi\, \vb{t}_{\beta} \vdot \vb{A}_1)$, 
where $\beta$ is the orbital index in the supercell, $V_0=200$~meV is the CDW amplitude, $\vb{t}_{\beta}$ is the sublattice vector of orbital $\beta$, and $\vb{A}_1$ is the first supercell lattice vector, defined by
\begin{equation}
    \mqty(\vb{A}_{1} & \vb{A}_{2} & \vb{A}_{3})^{\mathsf{T}} \equiv 
    M \mqty(\vb{a}_{1} & \vb{a}_{2} & \vb{a}_{3})^{\mathsf{T}},
    \qquad
    M = 
    \begin{pmatrix}
        4 & 2 & 0 \\
        0 & 1 & 0 \\
        0 & 0 & 1
    \end{pmatrix}.
\end{equation}

To compare with experiment, the surface Green’s function calculated in the $4\times1\times1$ supercell was unfolded back to the primitive Brillouin zone (pBZ)~\cite{Ku2010UnfoldingA,Popescu2012ExtractingA,Lee2013UnfoldingA,Miao2023TruncatedA}. The unfolded Green’s function is given by
\begin{equation}
    G^{\text{(unfold)}}(\bk, \omega) = X^{\dagger}\, G^{\text{(CDW)}}(\mathbf{K}, \omega)\, X,
\end{equation}
where $\bk$ is a momentum in the pBZ, $\mathbf{K}$ is the corresponding momentum in the supercell Brillouin zone, and $\vb{G}_n$ satisfies $\bk = \mathbf{K} + \vb{G}_n$. The unitary transformation has matrix elements 
$X_{\beta \alpha} = e^{i \vb{G}_n \vdot \vb{t}_{\beta}}\, \delta(\vb{t}_{\beta} - \vb{t}_{\alpha})$, 
with $\vb{t}_{\beta}$ ($\vb{t}_{\alpha})$ the sublattice vector in the CDW supercell (primitive cell). The unfolded QPI spectrum is then obtained by replacing $G_0 \to G_0^{\text{(unfold)}}$ in Eq.~(\ref{eqn:methods-qpi}).

\section*{Acknowledgement}
A.G.G. is supported by the National Science Foundation Graduate Research Fellowship Program under Grant No. DGE-2139841. Any opinions, findings, and conclusions or recommendations expressed in this material are those of the author(s) and do not necessarily reflect the views of the National Science Foundation.
E.H.d.S.N. acknowledges support from the National Science Foundation under grant number DMR-2034345 and Yale University startup funds.
J.D. acknowledges the computational resources provided by the Aalto Science-IT project. J.D. and P.T. were supported by the Finnish Centre of Excellence in Quantum Materials (QMAT). J.D., P.T., C.S., D.C. and C.F. were supported by a collaboration between The Kavli Foundation, Klaus Tschira Stiftung, and Kevin Wells, and by the Jane and Aatos Erkko Foundation, the Keele Foundation and the Magnus Ehrnrooth Foundation, as part of the SuperC collaboration. B.A.B and P.T. were supported by a grant from the Simons Foundation (SFI-MPS-NFS-00006741-01, B.A.B.; SFI-MPS-NFS-00006741-12, P.T.) in the Simons Collaboration on New Frontiers in Superconductivity. B.A.B was also supported by the Gordon and Betty Moore Foundation through Grant No. GBMF8685 towards the Princeton theory program, the Gordon and Betty Moore Foundation's EPiQS Initiative (Grant No. GBMF11070), the Office of Naval Research (ONR Grant No. N00014-20-1-2303), the Global Collaborative Network Grant at Princeton University, the Simons Investigator Grant No. 404513, the NSF-MERSEC (Grant No. MERSEC DMR 2011750), Princeton Catalysis Initiative, and the Schmidt Foundation at the Princeton University. Y. J., and H. H and partially BAB were supported by a European Research Council (ERC) under
the European Union’s Horizon 2020 research and innovation program (Grant Agreement No. 101020833). S.B-C. thanks the MINECO of Spain through the project PID2024-161503NB-C21.
This work was financially supported by the Deutsche Forschungsgemeinschaft (DFG) under SFB1143 (Project No. 247310070), the Würzburg-Dresden Cluster of Excellence on Complexity, Topology and Dynamics in Quantum Matter—ctd.qmat (EXC 2147, Project No. 390858490), and QUAST-FOR5249-449872909.

\section*{Author Contribution}
A.G.G, X.Y., P.K.N., K.S. and E.H.d.S.N. performed the STM measurements. A.G.G. performed the analysis of the experimental data under the supervision of E.H.d.S.N. and P.K.N..
J.D., Y.J., H.H., and B.A.B. provided theoretical guidance in the interpretation of the data.
J.D. conducted the \textit{ab initio} calculations, together with the help and supervision of Y.J., H.H., P.T., and B.A.B.. 
X.Y. performed the initial tight-binding calculation containing sublattice effects, and X.Y. and J.D. worked together in the final analysis of the \textit{ab initio} results. 
C.S., D.C., C.F. synthesized, characterized and provided the materials for this study. A.G.G., X.Y., J.D. and E.H.d.S.N. wrote the manuscript with the input from all other authors. 
E.H.d.S.N., S.B-C. and B.A.B. conceived of the experiments. 
E.H.d.S.N. was responsible for the overall planning and management of this work.

\section*{Competing Interests}
The data that support the findings of this study are available from the corresponding authors
upon request.

\newpage
\begin{figure}
    \centering
    \includegraphics[width=1\linewidth]{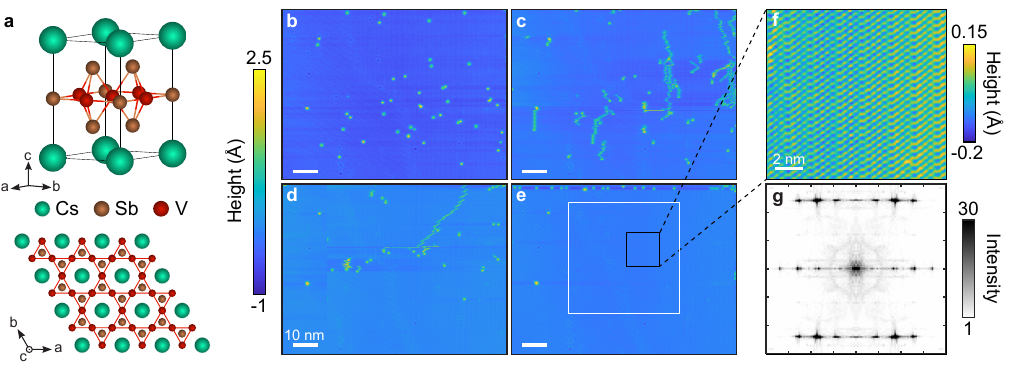}
    \caption{\textbf{Surface preparation for STS measurements.} \textbf{a}, The crystal structure of \CVS. \textbf{b}-\textbf{f}, Topographic images of the cleaved \CVS{} surface showing scattered Cs atoms before (\textbf{b}), during (\textbf{c}-\textbf{d}) and after (\textbf{e}-\textbf{f}) the atom sweeping process described in the text. The white square shows the area used for STS measurements. \textbf{g}, Fourier transform of the topography in \textbf{f}. STM topographic parameters: \textbf{b:} \(\mathrm{V_{bias}}\) = -50 mV, \(\mathrm{I_{set}}\) = 100 pA. \textbf{c:} \(\mathrm{V_{bias}}\) = -50 mV, \(\mathrm{I_{set}}\) = 350 pA. \textbf{d:} \(\mathrm{V_{bias}}\) = -50 mV, \(\mathrm{I_{set}}\) = 350 pA. \textbf{e:} \(\mathrm{V_{bias}}\) = -50 mV, \(\mathrm{I_{set}}\) = 100 pA. \textbf{f:} \(\mathrm{V_{bias}}\) = -50 mV, \(\mathrm{I_{set}}\) = 140 pA.} 
    \label{Fig1}
\end{figure}

\clearpage

\newpage
\begin{figure}
    \centering
    \includegraphics[width=1\linewidth]{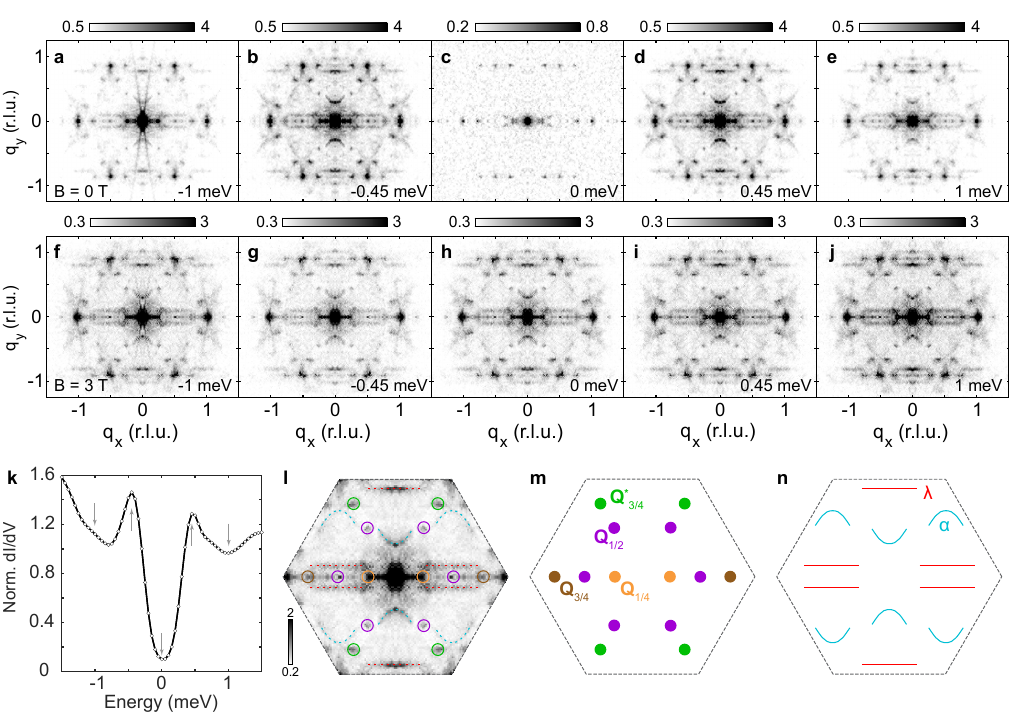}
    \caption{\textbf{High energy-sampling STS measurements.} \textbf{a}-\textbf{j}, Spectroscopic measurements at a range of energies from $-1$\,meV to $1$\,meV in the superconducting state (\textbf{a}-\textbf{e}) and when superconductivity is suppressed with a $3$\,T applied magnetic field (\textbf{f}-\textbf{j}). 
    {Note the distinct color scale in \textbf{c}, chosen to emphasize the Fourier pattern at $0\,\mathrm{meV}$, where the overall intensity is smaller.} 
    \textbf{k}, Spatially averaged d$I/\mathrm{d}V$ spectrum. Arrows highlight the energies displayed in (\textbf{a}-\textbf{j}). \textbf{l}, Energy-integrated $3$\,T spectroscopic measurements. Circles indicate the CDW features. \textbf{m}-\textbf{n}, Schematic representation of the CDW and QPI features, respectively. }
    \label{Fig2}
\end{figure}

\clearpage

\begin{figure}
    \centering
    \includegraphics[width=1\linewidth]{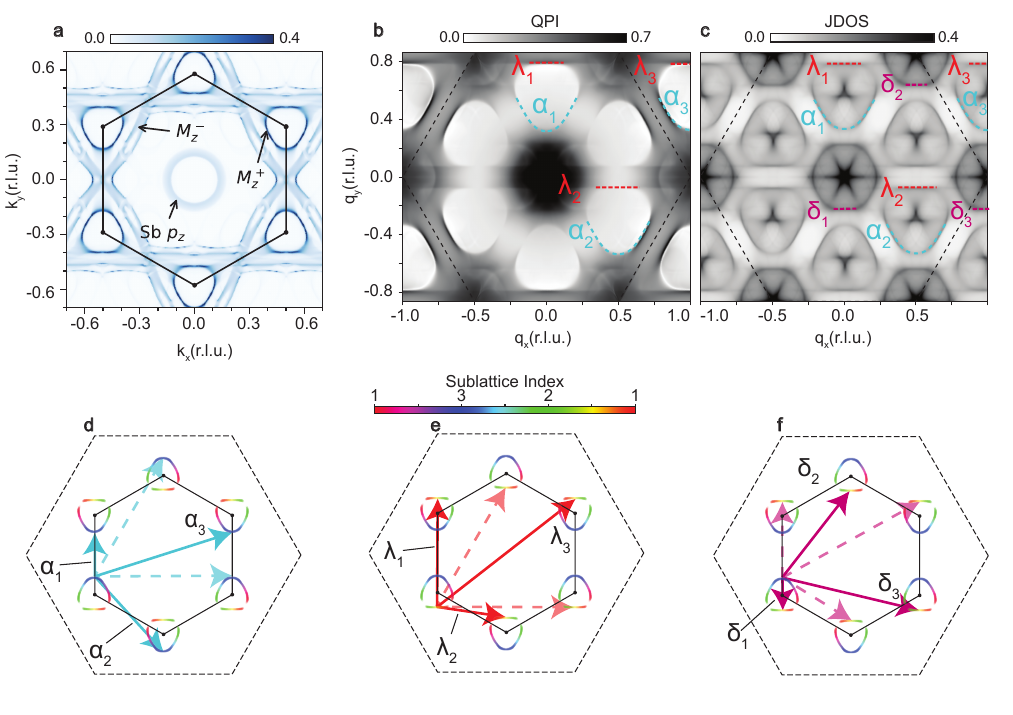}
    \caption{\textbf{$\bk \rightarrow \bq$ mapping and sublattice character.} \textbf{a}, First-principles surface Fermi surface, with the V-$M_{z}^{+}$, V-$M_{z}^{-}$, and Sb $p_z$ Fermi surfaces indicated by arrows. The $\bQ_{1/4}$ folding is included by adding a $4\times1\times1$ periodic on-site potential to the V kagome lattice, and the resulting spectrum is unfolded to the primitive Brillouin zone. \textbf{b-c}, QPI for the V-$M_{z}^{+}$ sector calculated via Green's-function and JDOS methods, respectively. \textbf{d-f}, V $M_z^+$ Fermi surface colored by sublattice weight (red, blue, and green). Solid arrows (cyan, red, and purple) indicate representative scattering processes that connect states with opposite group velocities and contribute to the QPI. Dashed arrows indicate processes suppressed by nearly parallel group velocities. The corresponding $\bq$-space features are indicated by dashed guides in \textbf{b,c}.
    }
    \label{Fig3}
\end{figure}

\clearpage

\begin{figure}
    \centering
    \includegraphics[width=1\linewidth]{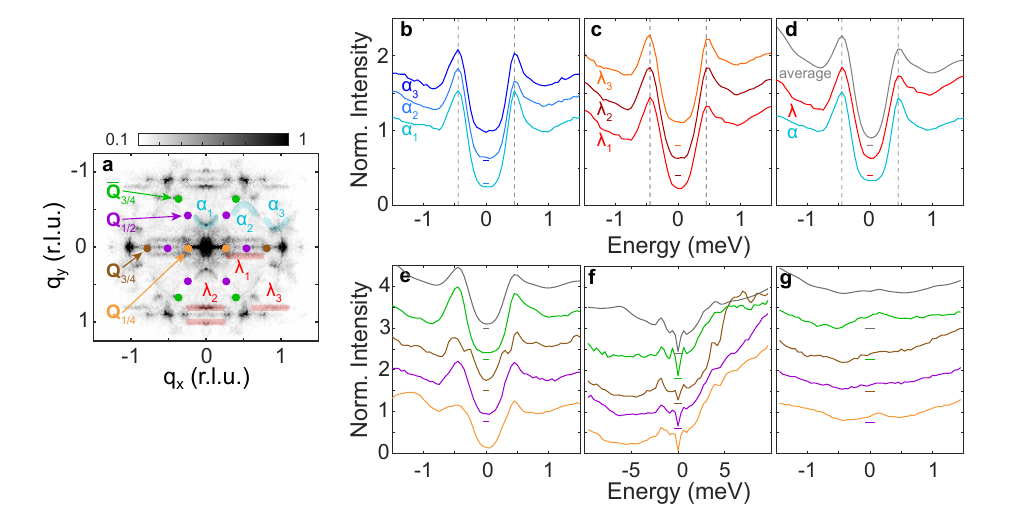}
    \caption{\textbf{Momentum resolved superconducting gap structure.}  \textbf{a}, Two-fold symmetrized Fourier transform of STS measurement taken on \CVS{} surface at bias voltage $V = 0.45$\,meV. \textbf{b}-\textbf{c}, Fourier intensity averaged over the $\alpha_{1-3}$ and $\lambda_{1-3}$ regions highlighted in \textbf{a}, respectively. \textbf{d}, Average of the $\alpha_{1-3}$ and $\lambda_{1-3}$ spectra in \textbf{b} and \textbf{c}, compared to the spatially averaged spectrum. \textbf{e}-\textbf{g}, Spectral intensity of the Fourier peak features marked in \textbf{a} at $B=0$\,T (\textbf{e},\textbf{f}) and $B=3$\,T (\textbf{g}). In \textbf{b}-\textbf{g}, all curves are normalized to their average over the displayed energy ranges.
    }
    \label{Fig4}
\end{figure}
\clearpage

\pagebreak

\end{document}